\def\comment#1{}

\newcommand{\beg}{\begin{eqnarray}}
\newcommand{\eee}{\end{eqnarray}}

\documentstyle[prl,aps,amsfonts,psfig,multicol,epsf]{revtex}
\def\cm#1{}

\begin{document}
\title{{ 
Aspects of topology of condensates and
knotted solitons 
in condensed matter systems
 }}
\author{
 Egor Babaev\thanks{Email: egor@teorfys.uu.se  \ \
 http://www.teorfys.uu.se/people/egor/   \ \
 \ \ Tel: +46-18-4717629  Fax +46-18-533180}}
\address{
Institute for Theoretical Physics, Uppsala University 
Box 803, S-75108 Uppsala, Sweden }
\comment{\\
}
\maketitle
\begin{abstract}
The knotted solitons introduced by Faddeev and Niemi is presently a subject of great interest in 
particle and mathematical physics. 
In this  Letter  we give a condensed matter interprtation of the 
recent results of Faddeev and Niemi . 
\end{abstract}
\newcommand{\la}{\label}
\newcommand{\Pfaff}{{\rm\, Pfaff}}
\newcommand{\G}{{\cal G}}
\newcommand{\cP}{{\cal P}}
\newcommand{\M}{{\cal M}}
\newcommand{\E}{{\cal E}}
\newcommand{\btd}{{\bigtriangledown}}
\newcommand{\W}{{\cal W}}
\newcommand{\X}{{\cal X}}
\renewcommand{\O}{{\cal O}}
\renewcommand{\d}{{\rm\, d}}
\newcommand{\bfi}{{\bf i}}
\newcommand{\e}{{\rm\, e}}
\newcommand{\bfx}{{\bf \vec x}}
\newcommand{\bfn}{{\bf \vec n}}
\newcommand{\bfE}{{\bf \vec E}}
\newcommand{\bfB}{{\bf \vec B}}
\newcommand{\bfv}{{\bf \vec v}}
\newcommand{\bfU}{{\bf \vec U}}
\newcommand{\bfp}{{\bf \vec p}}
\newcommand{\bfA}{{\bf \vec A}}
\newcommand{\non}{\nonumber}
\newcommand{\be}{\begin{equation}}
\newcommand{\ee}{\end{equation}}
\newcommand{\ba}{\begin{eqnarray}}
\newcommand{\ea}{\end{eqnarray}}
\newcommand{\bastar}{\begin{eqnarray*}}
\newcommand{\eastar}{\end{eqnarray*}}
\newcommand{\half}{{1 \over 2}}
\begin{multicols}{2}
\narrowtext

\comment{
External magnetic fields can penetrate into type II superconductors
through  cores of topological defects - Abrikosov vortices.
This phenomenon, because of its macroscopic quantum
origin, counter-intuitive behavior and simplicity 
of formal description was one of the most 
remarkable theoretical predictions in condensed 
matter physics. This concept was later generalized
to particle physics and evolved to a large 
and diverse part of theoretical  and mathematical physics.
 opological defects in superfluids is one the largest 
and most diverse part of condensed matter physics.}

Since the celebrated  work by Abrikosov \cite{aaa} on vortices in superconductors,
the field of the topological defects in superfluids evloved into 
one of the largest  and most diverse  parts of condensed matter physics \cite{vol}.

In this Letter we discuss possibilities of formation
in condensed matter systems  
 of the {\it knotted solitons} and their basic properties.
This principally new type of  topological 
excitations was introduced by Faddeev and Niemi
and was discussed first
in mathematical and high energy physics
 \cite{fadde}-\cite{plasma}. 
\comment{
An interesting 
circumstance  is that, as we show below, a Ginzburg-Landau
functional for a two-gap superconductor is equivalent to
 the same model \cite{fadde} 
that was  derived for QCD in the infrared limit \cite{nature}.
It is  an example of the universal character of 
the model  \cite{fadde} that  
appears to describe  
a wide range of physical systems
despite, on first glance, their
 different physical origins.}
It appears  that condensed matter physics 
is a field with 
an especiailly  wide range of possible applications 
 for the of the concept of knotted solitons. 
\comment{ such as condensate 
desity profile, the notion  of the  knotted soliton core
(which is principally different from the ordinary vortex core),
and geometry of supercurrents.
Structural  properties 
of a knot solition in a superfluid being  principally different 
from the structure of ordinary vortex thus open 
a wide range of questions associated with formation and 
interaction of these defects.}

A formation of a knot soliton requires  
a system with more degrees of freedom 
and different topology than  ordinary 
BCS superconductors. In particular, as discussed
in the  abovementioned papers by Faddeev and Niemi
the knot solitons may form in a system of
two  charged scalar fields.
In condensed matter physics the system 
of two charged 
 complex scalar fields were discussed earlier 
in context of two-band  superconductivity 
\cite{tg}.
In two gap superconductors a
Fermi surface passes through two bands thus 
giving a theoretical possibility 
to a formation of two superconductive condensates.
Presently, there are also ongoing discussions of the possibile  
coexistence of two condensates  in High-Temperature
Superconductors. 
Another  condensed matter system 
which  theoretically   allows  two condensates  is the
 liquid metallic hydrogen and deuterium.
Liquid metallic hydrogen should allow coexistentce of superconductivity 
of  electronic and protonic Cooper pairs , as it was investigated
by
Moulopoulos and Ashcroft \cite{ashc1}. In a liquid metallic 
deuterium a superfluidity of deutrons may coexist with 
superconductivity of electronic
Cooper pairs \cite{ashc2} (see also \cite{ashc3}) .

A system of two coupled through magentic field
charged condensates can be described by means 
of a two-flavour (denoted by $i=1,2$) Ginzburg-Landau functional:
\beg
F = \sum_{i=1,2}\int d^3x \ \biggl[ \frac{1}{4m_1} | (\partial_k +
i 2 e A_k) \Psi_1 |^2 + \nonumber \\
+ \frac{1}{4m_2} | (\partial_k -
i 2 e A_k) \Psi_2 |^2
 + b_i|\Psi_i|^2+ \frac{c_i}{2}|\Psi_i|^4
+ \frac{H^2}{8\pi}
\biggr]
\la{act}
\eee
We consider a general case when condensates are characterized
by different effective masses $m_i$, densities $<\Psi_i>$
 and coherence lengthes $\xi_i$. For this system 
we derive an effective Faddeev model following 
to  Faddeev-Niemi method.

  The two condensates are coupled through an 
electromagnetic field. Variation with respect to $A_k$
gives the following expression for the magnetic 
field in  such a system:
\beg
&&\frac{1}{2}\mbox{rot} {\bf B}= \sum_i 
\frac{2e^2}{m_i} {\bf A} |\Psi_i|^2
-\frac{e}{2m_1}
\left\{\Psi_1^*\partial \Psi_1-
\Psi_1 \partial \Psi_1^*\right\}
 \nonumber \\
&&
+\frac{e}{2m_2}
\left\{\Psi_2^*\partial \Psi_2-
\Psi_2 \partial \Psi_2^*\right\}.
\la{A}
\eee

Let us rewrite the gradient term in the following form:
\beg
E  =  \int d^3 x \biggl[ \
\frac{\Xi}{4 } \biggl\{ ~ \sin^2 \gamma \, |
(\partial_k + i 2 e A_k)
f_1 |^2  +  \nonumber \\ +\cos^2  \gamma \, | (\partial_k -
i 2 e A_k) f_2 |^2 ~ \biggr\}
\eee
\comment{
+  \frac{1}{2} B_{i}^2 \ + \ g \bigl( {\psi_e}^* \psi_e - \nonumber \\
{\psi_i}^* \psi_i\bigr)^2 \ \biggr]
\la{ene}
\ee}
Here $f_i = |\Psi_i|/{\bar \Psi_i}$, where 
$\Xi^{-1} = (m_1/{\bar \Psi_1^2})  \sin^2  \gamma 
= (m_2/{\bar \Psi_2^2})  \cos^2  \gamma$ and
${\bar \Psi_i}$ stands  
for the average value of $\Psi_i$.

\comment{
We start by observing that the vector
potential $A_k$ enters at most quadratically. Consequently
it can be eliminated: We vary (\ref{ene}) {\it
w.r.t.} $A_k$ and get
\[
A_k \ = \  \frac{1}{2 e}\cdot \frac{1}{\sin^2\! \alpha \,
|\psi_e|^2 + \cos^2 \! \alpha \,  |\psi_i|^2}
\biggl[ \, i\sin^2 \! \alpha \cdot
({\psi_e}^* \partial_k \psi_e - \partial_k {\psi_e}^* \psi_e) \]
\be
-i \cos^2 \! \alpha \cdot ( {\psi_i}^* \partial_k \psi_i -
\partial_k {\psi_i}^* \psi_i ) \ - \
\frac{2\mu}{e} \cdot \epsilon_{kij}\partial_i
B_{j} \biggr]
\la{A}
\ee
which determines $A_k$ in terms of an iterative
gradient expansion, in powers of derivatives in the
charged fields.}
We parametrise the variables $(f_1,f_2)$  as:
\beg
&&\left(  f_1 ({\bf x})  ,  f_2({\bf x})  \right)  = \nonumber \\
&&\sigma({\bf x}) \cdot \left(  \cos\gamma \sin \frac{\theta({\bf x})}{2} \cdot
 e^{i \varphi_1({\bf x})}  , 
\sin\gamma \cos \frac{\theta({\bf x})}{2} \cdot e^{i\varphi_2({\bf x})}  \right)
\la{Psi}
\eee
where the variable $\sigma({\bf x})$ is chosen so that, for 
 uniform consendates $|\Psi_i ({\bf  x_\infty})| = {\bar \Psi_i}$ it
assumes the value $\sigma^* =2/\cos(2\gamma)$.
Following to Faddeev and Niemi method
 we are using equations of motion to eliminate  gauge field and reexpress the free energy (\ref{act})
to next-to-leading leading order in the  gradients of condensates.
That is,
we determine $A_k$ from (\ref{A}) in the
variables $f_{1,2}$ and  substitute the result
in (\ref{act}).  As the next step we
define a three-component
unit vector \cite{nature}
\beg
\bfn  = ( \cos(\varphi_1+\varphi_2) \sin
\theta  \, ,  \, \sin(\varphi_1 + \varphi_2)\sin \theta \, , \,
\cos \theta )
\la{v}
\eee
and eliminate the $f_{1,2}$ fields
in favour of the new variables ${\bfn}$ and $\sigma$. 
Thus we arrive to an effective Faddeev model 
for two condesates of unequal masses and densities:
\beg
 &&F=\int d^3x\biggl\{\frac{1}{4}\left[ \frac{m_1}{\bar \Psi_1^2} 
+\frac{m_2}{\bar \Psi_2^2}\right]^{-1}(\partial_k \sigma)^2+ 
\nonumber \\
&& \frac{ \sigma^2 }{16}
\left[ \frac{m_1}{\bar \Psi_1^2} 
+\frac{m_2}{\bar \Psi_2^2}\right]^{-1}
|\partial_k \bfn |^2+
\frac{1}{128e^2}(\bfn \cdot \partial_i
\bfn \times \partial_j\bfn
)^2 \, + \nonumber \\
&&\frac{\sigma^2}{2} \biggl(
\biggl[ {b_1 {\bar \Psi_1^2}\cos^2\gamma+b_2{\bar \Psi_2^2}\sin^2\gamma} +
\nonumber \\
&&\frac{ \sigma^2}{4} (c_1{\bar \Psi_1^4}\cos^4\gamma
 +c_2{\bar \Psi_2^4}\sin^4\gamma)  \biggr] -
\nonumber \\
&&\biggl[ {b_1{\bar \Psi_1^2}\cos^2\gamma-b_2{\bar \Psi_2^2}\sin^2\gamma} +
\nonumber \\
&& \frac{\sigma^2}{2}(c_1{\bar \Psi_1^4}\cos^4\gamma
-c_2{\bar \Psi_2^4}\sin^4\gamma) \biggr] n_3+
\nonumber \\
&&\frac{ \sigma^2}{4} (c_1{\bar \Psi_1^4}\cos^4\gamma
+c_2{\bar \Psi_2^4}\sin^4\gamma) \ n_3^2 \biggr)
  \biggr\}
\la{fadd}
\eee
This model has similar structure with the model that was considered in \cite{nature}
\be
H = \varepsilon \int d^3x \ \biggl[
\ \alpha \cdot |\partial_k \bfn |^2  \ + \ \frac{1}{4e^2}
(\bfn \cdot \partial_i \bfn \times \partial_j\bfn
)^2 \biggr]
\la{fad2}
\ee
A system characterized by such a Hamiltonian admits
stable knotted solitons \cite{nature}. 
\comment{This feature
makes this object stable against shirnkage 
thus being in contrast to a closed Abrikosov vortex
loop which energy is proportional to its length 
that makes this object instable.}
In terms of the vector $\bfn$ the knot soliton 
in the Faddeev model (\ref{fad2})
 has the following structure:
At large distances away from the knot,
 the unit vector $\bfn ({\bf x} )$ should 
approach   
the constant value $\bfn_0$. Let us choose this asymptotic value
to be the south pole
$\bfn_0=(0,0,-1)$. The pre-image of the north pole $\bfn_c=(0,0,1)$
is a closed curve,  associated with  the core of the closed vortex
and may form a knot or a loop.
The unit vector $\bfn$
defines a point on $S^2$  and thus 
 $\bfn ({\bf x})$ defines  a map from the compactified 
$R^3 \propto S^3 \rightarrow S^2$. Such mappings
fall into nontrivial homotopy 
classes $\pi_3(S^2)\simeq Z$ that can be characterized
by a  Hopf invariant.
\comment{One can
 introduce a closed two-form $F=(d \bfn \wedge \bfn, \bfn)$
on the target $S^2$. Since $H_2(S^3)=0$ its preimage on the base $S^3$
is exact $F_\star=d A_\star$ and  the Hopf invariant $Q_H$ coincides
with the three dimensional Chern-Simons term \cite{fadde,nature}
\be
Q_H=c \frac{1}{4\pi^2} \int_{R^3} F \wedge A.
\label{qq}
\ee
The estimate \cite{vak} 
defines lower bound of the energy for this model
as fractional power  of the Hopf invariant:
\be
E \geq c |Q_H|^{3/4}
\ee 
\comment{\it This estimate states that the first two
terms in (\ref{fad2}) are bounded from below by the 
fractional power $|Q_H|^{3/4}$ of the Hopf invariant. 
Even though we do not expect that in the case
of (\ref{fadd}) this lower bound estimate 
remains valid as such,
we nevertheless conclude that when $Q_H \not=0$ the energy
(\ref{fadd}) admits a nontrivial lower bound;}
}
The  variables of (\ref{v})
 describe a helical structure, with  the
Hopf invariant being given by \cite{nature}: 
\beg
&&Q_H = \  -(e^2 16\pi^2)^{-1} \int d^3x  \bfB \cdot \bfA  =
\nonumber \\
&& \int d^3x  \nabla  \cos \theta \cdot
{\nabla \varphi_1}/{2\pi} \times
{\nabla \varphi_2}/{2\pi}  =  \Delta \varphi_1\cdot \Delta
\varphi_2\eee
Here $\Delta \varphi_1$ and $\Delta \varphi_2$ denote the
($2\pi$) changes  in the phases of the wave functions 
of condensates along a path covering
the closed/knotted  vortex tube
 once in the toroidal and poloidal
directions over a surface with constant $\theta \in (0,\pi)$.
In the case of  two-condensate model 
(\ref{act}) there is no 
exact $O(3)$ symmetry as it is also explicitly  seen from  (\ref{fadd}).
That is, 
 the mass terms for  third component $n_3$ of the unit vector $\bfn$
breaks the O(3)-symmetry. Whereas $(n_1,n_2)$ components
of the unit vector $\bfn$ are related to the phases of the 
condensates and thus this symmetry is exact, in contrast 
the $n_3$ component is connected with a degree of freedom 
associated with massive modules of the scalar fields - thus 
the $n_3$ componet is massive and has a preferable value.
In the model under the consideration in this paper 
the energetically preferable value for $n_3$ is:
\be
 n_3^* = \left[\frac{\bar \Psi_1^2}{m_1} -\frac{\bar \Psi_2^2}{m_2}\right] \cdot
\left[ \frac{ \bar \Psi_1^2}{m_1} + \frac{\bar \Psi_2^2}{m_2}\right]^{-1}
\label{nstar}
\ee
Albeit one of the components of the
unit vector $\bfn$
is massive, it does not affect the stabilty of the knot solitions 
in the action (\ref{fadd}) which are protected against shrinkage by the 
Faddeev term [third term in (\ref{fadd}) ]. However
the mass and preferable value for $n_3$ should affects the 
geometry of the knotted solitons.
 
In conclusion we discussed realisations of knotted solitions in a general 
system of two condensates giving a condensed matter interpretation of the 
Faddeev and Niemi results
\cite{nature,plasma}.
A remarkable 
circumstance is that such a condensed matter system which 
may describe e.g. the  coexistent 
electronic and protonic superconductivity in a liquid metallic hydrogen 
\cite{ashc1}  is described 
by an effective model introduced in  \cite{fadde}
that was derived before as an effective model
for QCD in the infrared limit \cite{nature}.
Thus it is a manifestation of the relevance 
of this model for a exceptionally  wide variety of physical problems 
and close relation of the problems 
of condensed matter and particle 
physics \cite{ppg}. A number of exotic 
features of these
topological defects in a general 
two-condensate system  opens an exceptionally 
 wide range of interesting questions associated with properties, formation
and interaction of these defects.
\comment{
Morover albeit the Faddeev term in effective  action 
appears only in a charged system, so , in contrast 
a similar topological defect 
in a neutral system  is not stable, still it may appear  in dymanic processes
thus being of relevanmce to multicomponent neutral Bose condensates
which are presently a subject of great interest. In a neutral system it would 
be very similar to the defects 
characterised by nontrivial Hopf invariant discussed earlier in 
\cite{vo2} which exception that in our case the component $n_3$
is amssive and there is no exact O(3) }
\comment{
A separate question, that we do not discuss
in this paper, is the possibility of an experimental
setup that would allow one to create and observe a knotted soliton 
in a two-band superconductor (e.g. in transition metals).
From an experimental
point of view, an interesting 
circumstance is that a two-band superconductor
 may serve as a ``testing laboratory"
for QCD  in sense that the possibility
of an experimental investigation of knotted solitons,
 along with ongoing numerical simulations, may  help 
to  illuminate the infrared  properties of QCD.}
 
The author is grateful to Prof. L.D. Faddeev, Prof. A.J. Niemi, Prof. G. E. Volovik,
Prof. N.W. Ashcroft, Dr. V. Cheianov, 
Dr. S. Ktitorov
and Prof. A.J. Leggett    for
  discussions or useful comments. 

{\bf Note added:} Recently an exact mapping  of Ginzburg-Landau model 
to a version of the Faddeev model without performing a derivative expansion 
was discussed in \cite{added}

\end{multicols}

\begin{thebibliography}{99}
\bibitem{vol}
In particular   more complicated  topological defects   in ${}^3 He$
chracterized by a nontrivial Hopf 
invariant were discussed in  \cite{vo2} (for an excellent 
review and citations see also \cite{vol3}).
\end{thebibliography}
\end{document}